\documentclass[href,cits,proceedings]{PoS}
\usepackage{amsmath}
\usepackage{mathbbol}
\newcommand{\dslash}{\! \not \!\! D}

\title{Disconnected contributions to hadronic structure}

\ShortTitle{Disconnected contributions to hadronic structure}

\author{\speaker{Sara Collins}, Gunnar Bali,  Andrea Nobile,
    Andreas Sch\"afer\\
  Institut f\"ur Theoretische Physik, Universit\"at Regensburg,\\
  93040 Regensburg, Germany\\
  E-mail: \email{sara.collins@physik.uni-regensburg.de}\\
\email{gunnar.bali@physik.uni-regensburg.de}\\
\email{andrea.nobile@physik.uni-regensburg.de}\\
\email{andreas.schaefer@physik.uni-regensburg.de}}

\author{Yoshifumi Nakamura\\
Center for Computational Sciences, University of Tsukuba,\\
Tsukuba, Ibaraki 305-8577, Japan\\
        E-mail: \email{yoshi@ccs.tsukuba.ac.jp}}

\author{James Zanotti\\
        School of Physics, University of Edinburgh,\\ Edinburgh EH9 3JZ, UK\\
        E-mail: \email{jzanotti@ph.ed.ac.uk}}

\author{
\rm\centering (QCDSF Collaboration)}

\abstract{We present an update of an on-going project to determine the
disconnected contributions to hadronic structure, specifically, the
scalar matrix element, $\langle N|\bar{q}q|N\rangle$, and the quark
contribution to the spin of the nucleon $\Delta q=\langle
N|\bar{q}\gamma_\mu\gamma_5{q}|N\rangle/m_N$.}

\FullConference{The XXVIII International Symposium on Lattice Field Theory, Lattice2010\\
		June 14-19, 2010\\
		Villasimius, Italy}

\begin{document}

\section{Introduction}
In the last few years there has been an upsurge in interest in
calculating the scalar matrix element on the
lattice~\cite{Ohki:2009mt,Babich:2009rq,Freeman:2009pu,ramos,jung}
either directly, by calculating the corresponding connected and disconnected terms,
or indirectly, via the Feynman-Hellman theorem. High statistics for
dynamical simulations mean that reasonable signals can be obtained for
disconnected terms and similarly the small statistical uncertainty on
nucleon mass as a function of the quark masses enable reasonable
fits to be made. Ideally, the results of both approaches should agree.

Such calculations have also become particularly timely since the advent
of the LHC because
the scalar coupling $f_{T_q}=m_q\langle N|\bar{q}{q}|N\rangle/m_N$
determines the fraction
of the proton mass $m_N$ that is carried by quarks of flavour $q$.
The strength of the coupling of the Standard
Model (SM) Higgs boson or of any similar scalar particle to the proton
is mainly determined by $\sum_q f_{T_q}$ for $q\in\{u,d,s\}$.
Therefore, an accurate calculation of
these quantities will help to increase the precision of SM phenomenology
and to shed light on non-SM processes.

The spin of the nucleon can be
decomposed into a quark spin contribution $\Delta\Sigma=\Delta u +
\Delta d+ \Delta s+\ldots$, a quark angular momentum contribution
$L_q$ and a gluonic contribution (spin and angular momentum) $\Delta
G$:
\begin{equation}
\frac12=\frac12 \Delta\Sigma+L_q+\Delta G\,.
\end{equation}
Experimentally, $\Delta s$ is not well determined: 
HERMES obtained~\cite{Airapetian:2007mh} $\Delta
s=-0.085(13)(8)(9)$ in the $\overline{MS}$ scheme. However, the signal
is dominated by contributions in the small $x$ region where models are
used to extrapolate from the experimental results obtained at larger $x$.

In these proceedings we present an update of an on-going project to
calculate $f_{T_q}$ and $\Delta q$. In particular, the following
improvements have been implemented since Lattice
2009~\cite{Bali:2009dz}:
\begin{itemize}
\item Statistics on the $24^3\times 48$ and $32^3\times 64$ lattices
  have been significantly increased, by factors of roughly two and
  three, respectively, using the SFB/TR55 QPACE
  computers~\cite{nobile,yoshifumi}~(details are given in the next section).
\item An additional volume of $40^3\times 64$ has been analyzed.
\end{itemize}
The analysis is not yet finalized and we plan to further increase the statistics
for the larger two volumes. The results presented here are preliminary.

\section{Simulation details}
The simulations were performed on $n_{\mathrm f}=2$ 
configurations of nonperturbatively improved clover fermions with
Wilson gauge action at $\beta=5.29$ and $\kappa_{\mathrm{sea}}=0.13632$.
Details of the volumes and the number of trajectories analyzed are given
in table~\ref{configs}. The pseudoscalar mass corresponding to this
$\kappa_{\mathrm{sea}}$ value is around $270$~MeV, using an inverse lattice spacing
of $2.59$~GeV determined from $r_0(\beta,\kappa)=0.467$~fm.

\TABULAR{|c|ccc|}{\hline
\multicolumn{4}{|c|}{$\beta=5.29$, $n_{\mathrm f}=2$, $\kappa_{\mathrm{sea}}=0.13632$}\\\hline
Volume & $24^3\times48$ & $32^3\times64$ & $40^3\times 64$\\\hline
\#~traj. & $\approx 2000$ & $\approx 1600$ & $\approx 730$\\
$m_{\mathrm{PS}} L$ &2.6 &3.4 &4.3 \\\hline}{Details of the configurations used.
\label{configs}}

\begin{figure}
\centerline{
\includegraphics[height=.2\textwidth]{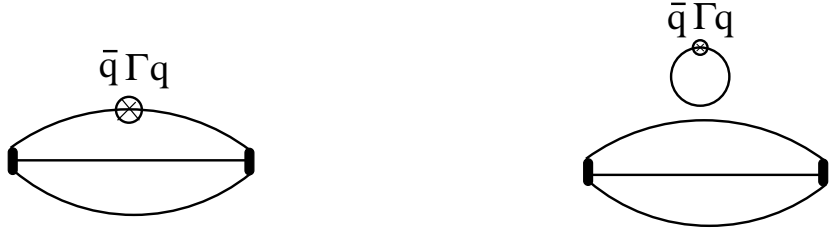}}
\caption{The connected~(left) and disconnected~(right) diagrams
  associated with the scalar~($\Gamma=\mathbb{1}$) and axial vector matrix elements~($\Gamma=\gamma_5\gamma_i$).}
\label{diags}
\end{figure}

The scalar and axial-vector matrix elements we are interested in are
extracted on the lattice from the three-point functions corresponding to
the diagrams given in fig.~\ref{diags}. The axial-vector matrix
element is related to $\Delta q$ through,
\begin{eqnarray}
\langle N,s|\bar{q}\gamma_{\mu}\gamma_5 q|N,s\rangle & = & 2m_N s_{\mu} \frac{\Delta q}{2},
\end{eqnarray}
in Minkowski space notation, where $m_N$ is the nucleon mass and
$s_\mu$ its spin~($s_\mu^2=-1$). The connected three-point functions
were calculated separately by QCDSF, details of which can be found in
refs.~\cite{qcdsf,Gockeler:2009pe}. The study presented here is only concerned
with the disconnected terms. For $\Delta s$ and
$\langle N|\bar{s}s|N\rangle$ only these terms contribute.

We varied the quark mass of the current insertion~(disconnected loop),
as well as the mass of the valence quarks in the nucleon. In the
following, we denote the $\kappa$ value corresponding to the quark
loop and the valence quarks in the nucleon as $\kappa_{\mathrm{loop}}$
and $\kappa_{\mathrm{val}}$, respectively. All combinations
$\kappa_{\mathrm{loop}}, \kappa_{\mathrm{val}}\in\{0.13550, 0.13609,
0.13632\}$ were used.  These values correspond to the pseudoscalar masses,
$m_{\mathrm{PS}}\approx 690$, $440$ and $270$~MeV, respectively.
The heaviest $\kappa_{\mathrm{val}}=0.13550$ roughly corresponds 
to the strange quark mass.

The disconnected contributions to $\Delta q$ and $\langle N|\bar{q}q|N\rangle$ were extracted from the
ratios of three-point functions to two-point functions~(at zero
momentum),
\begin{equation}
\label{eq:rati}
R^{\mathrm{dis}}(t_{\mathrm f},t,t_{\mathrm i}) = 
-
\frac{\mathrm{Re}\,\left\langle\Gamma_{\mathrm{2pt}}^{\alpha\beta}C^{\beta\alpha}_{\mathrm{2pt}}(t_{\mathrm f},t_{\mathrm i}) \sum_{\mathbf{x}}\mathrm{Tr}\,(M^{-1}(\mathbf{x},t;\mathbf{x},t)\Gamma_{\mathrm{loop}})\right\rangle}{\left\langle \Gamma_{\mathrm{unpol}}^{\alpha\beta} C^{\beta\alpha}_{\mathrm{2pt}}(t_{\mathrm f},t_{\mathrm i})\right\rangle}\,,
\end{equation}
where the nucleon source and sink are at $t_{\mathrm i}$ and $t_{\mathrm f}$
respectively, and the current is inserted at $t$.  The three-point
function is simply the combination of the nucleon two-point function,
$C_{\mathrm{2pt}}(t_{\mathrm f},t_{\mathrm i})$, and the disconnected loop,
$ \sum_{\mathbf{x}}\mathrm{Tr}\,[M^{-1}(\mathbf{x},t;\mathbf{x},t)\Gamma_{\mathrm{loop}}]$. For the scalar matrix element we used,
$\Gamma_{\mathrm{2pt}}=\Gamma_{\mathrm{unpol}}:=(\mathbb{1}+\gamma_4)/2$ and $\Gamma_{\mathrm{loop}}=\mathbb{1}$. For $\Delta q$ we calculated the difference between two
polarizations: $\Gamma_{\mathrm{2pt}}= \gamma_j\gamma_5\Gamma_{\mathrm{unpol}}$
and $\Gamma_{\mathrm{loop}}=\gamma_j\gamma_5$, where we average over all
three possible $j$-orientations.

\begin{figure}
\centerline{
\rotatebox{270}{\includegraphics[height=.48\textwidth,clip]{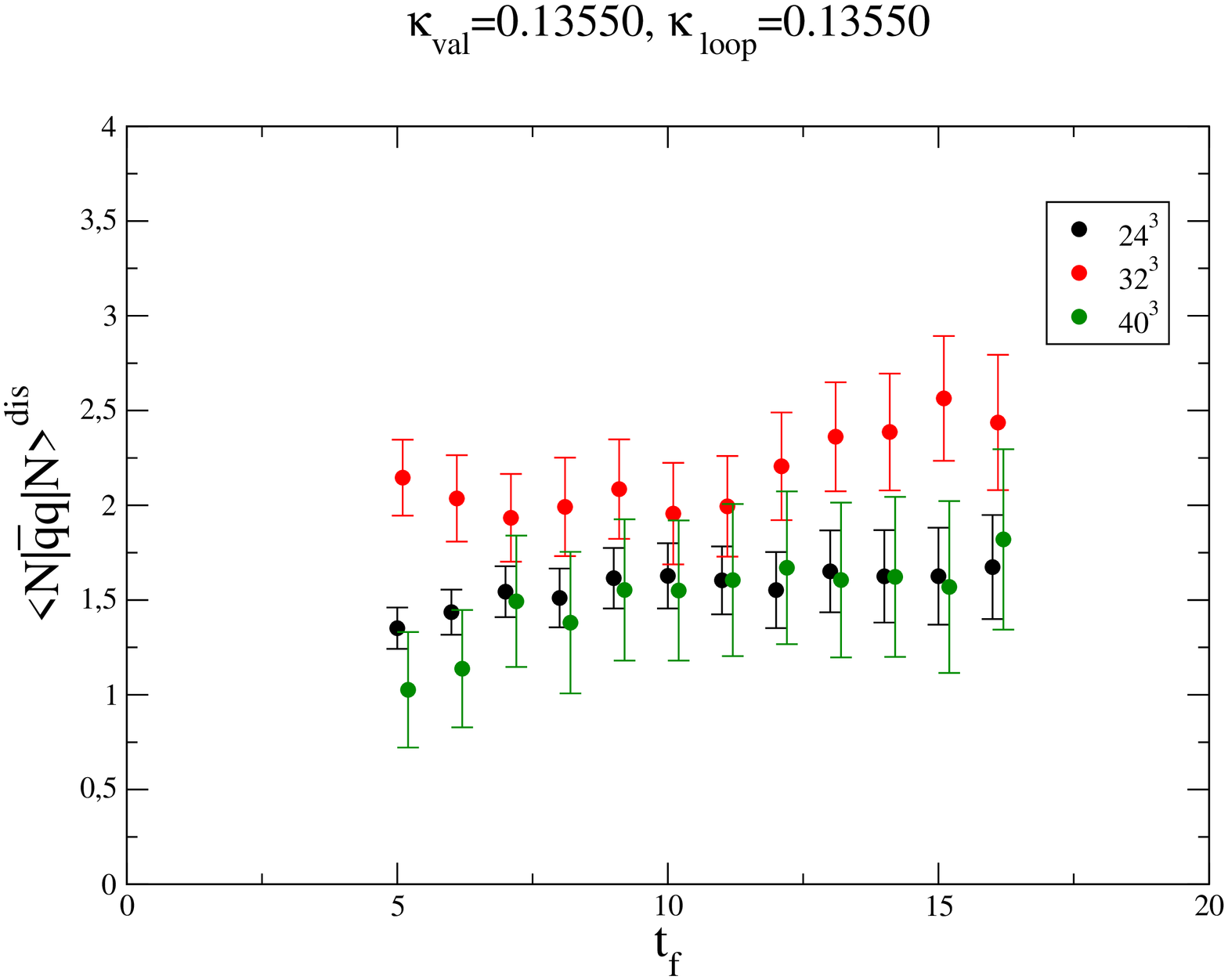}}\hspace*{.03\textwidth}
\rotatebox{270}{\includegraphics[height=.48\textwidth,clip]{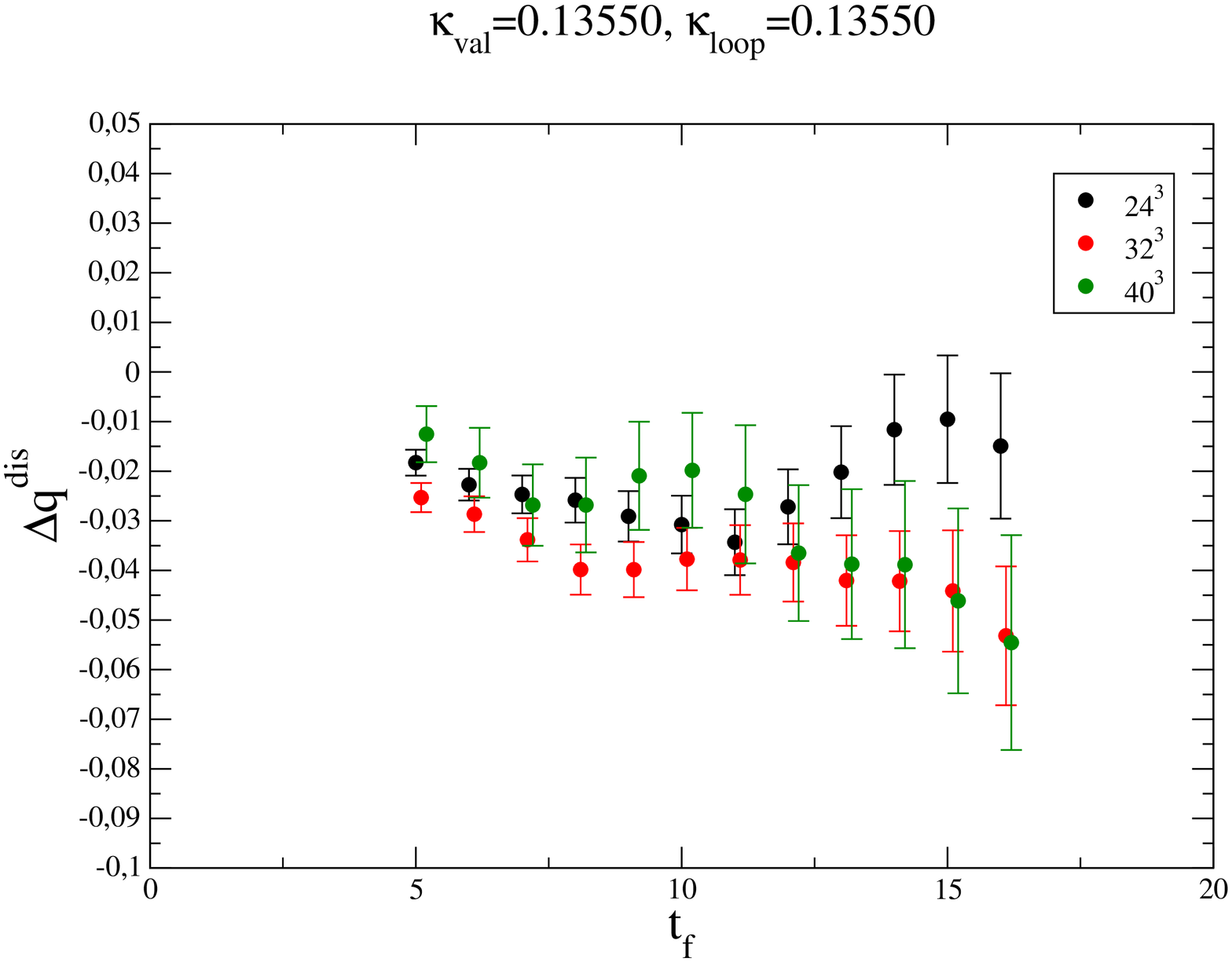}}}
\caption{The results for $\langle N|\bar{q}q|N\rangle^{\mathrm{dis}}$ and
  $\Delta q^{\mathrm{dis}}$ extracted from the corresponding $R^{\mathrm{dis}}$ as a
  function of the sink timeslice, $t_{\mathrm f}$, for all three volumes used and
  the heaviest $\kappa_{\mathrm{val}}=\kappa_{\mathrm{loop}}=0.13550$ combination.}
\label{rdis}
\end{figure}

In the limit of large times, $t_{\mathrm f}\gg t\gg t_{\mathrm i}$, depending
on the $\Gamma$-combination used, $R^{\mathrm{dis}}$
will either approach the disconnected axial matrix element $\Delta q^{\mathrm{dis}}$ or
the disconnected scalar matrix element $\langle N|\bar{q}q|N\rangle^{\mathrm{dis}}$ (once the
vacuum contribution is subtracted). However, the statistical noise
increases rapidly with increasing $t-t_{\mathrm i}$ and this time
difference needs to be minimized, using smeared sources and sinks for
the nucleon, in order to obtain a reasonable signal. A smearing study
on a limited number of configurations indicated that the nucleon
plateaued around $t\ge 4a\approx 0.3\,$fm and we chose to insert the
current at this timeslice~($t_{\mathrm i}=0$).  However, the higher statistics
now available mean the ground state dominates around $t\ge 6a$ only. At zero
momentum, the excited state contribution to $R^{\mathrm{dis}}$ is governed by
the time difference, $t_{\mathrm f}-t_{\mathrm i}$, and we must be careful to
choose $t_{\mathrm f}$ large enough.

In fig.~\ref{rdis} we show the results for $R^{\mathrm{dis}}$ as a function
of $t_{\mathrm f}>t$ for all three volumes studied and the heaviest
$\kappa_{\mathrm{val}}=\kappa_{\mathrm{loop}}=0.13550$ combination. So far
we have chosen $t_{\mathrm f}=8a$ or $9a$ based on the quality of the
plateau within given statistical errors. This depends on the
observable and lattice volume. 
Once final statistics are reached we will fit
the three- and two-point functions within
$R^{\mathrm{dis}}$ of eq.~(\ref{eq:rati})
separately, as functions
of $t_{\mathrm f}$, in order to extract the asymptotic values. 

The disconnected loop,
$ \sum_{\mathbf{x}}\mathrm{Tr}\,[M^{-1}(\mathbf{x},t;\mathbf{x},t)\Gamma_{\mathrm{loop}}]$, was calculated
using stochastic estimates, together with several noise reduction
techniques:
\begin{itemize}
\item Partitioning~\cite{Bernardson:1993yg,Wilcox:1999ab}: the stochastic source has
  support on eight timeslices.  Additional two-point functions were
  generated for four time separated source points on each
  configuration. The forward and backward propagation from these
  four source points was combined with the loop to give us eight
  measurements of $R^{\mathrm{dis}}$ per configuration.
\item Hopping parameter expansion~\cite{Thron:1997iy}: for the clover action the first two
  terms in the expansion of the disconnected loop,
\begin{equation}
\mathrm{Tr}(M^{-1}\Gamma_{\mathrm{loop}})=
2\kappa\mathrm{Tr}[(\mathbb{1}-\kappa\dslash)^{-1}\Gamma_{\mathrm{loop}}]=
\mathrm{Tr}[(2\kappa\mathbb{1}+2\kappa^2\dslash+\kappa^2\dslash^2M^{-1})\Gamma_{\mathrm{loop}}]\,,
\end{equation}
vanish and hence only contribute to the noise.
This means
$\mathrm{Tr}\,[\kappa^2\dslash^2M^{-1}(\mathbf{x},t;\mathbf{x},t)\Gamma_{\mathrm{loop}}]$ can be
used as an improved estimate of the loop. 
(In the case of $\Gamma_{\mathrm{loop}}=\mathbb{1}$ the non-vanishing
first term $\sum_{\mathbf{x}}2\kappa\mathrm{Tr}\,\mathbb{1}=24\kappa L^3$ can easily be corrected for.)
\item Truncated solver method~\cite{Bali:2009hu}: $730$ conjugate
  gradient solves were used, where the solver was truncated after $40$
  iterations. $50$ BiCGStab solves running to full convergence were
  generated to correct for the truncation error.
\end{itemize}
The noise reduction techniques other than time partitioning are only
necessary for determining $\Delta q$; for the scalar matrix element
the gauge noise dominates.

\section{The scalar matrix element: $f_{T_s}$ and $m_q\langle N|\bar{q}q|N\rangle^{\mathrm{dis}}$}
\begin{figure}
\centerline{
\rotatebox{270}{\includegraphics[height=.48\textwidth,clip]{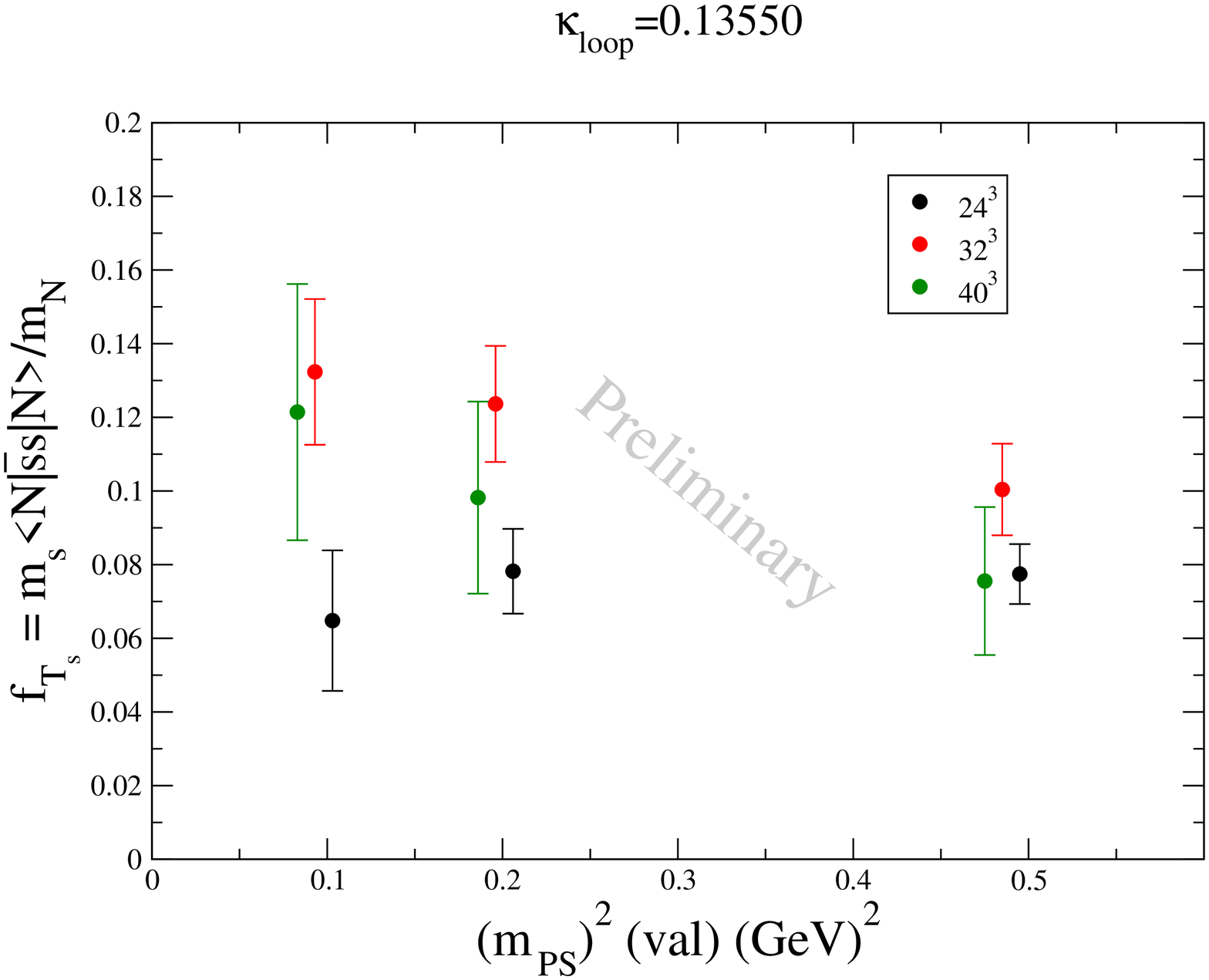}}\hspace*{.03\textwidth}
\rotatebox{270}{\includegraphics[height=.48\textwidth,clip]{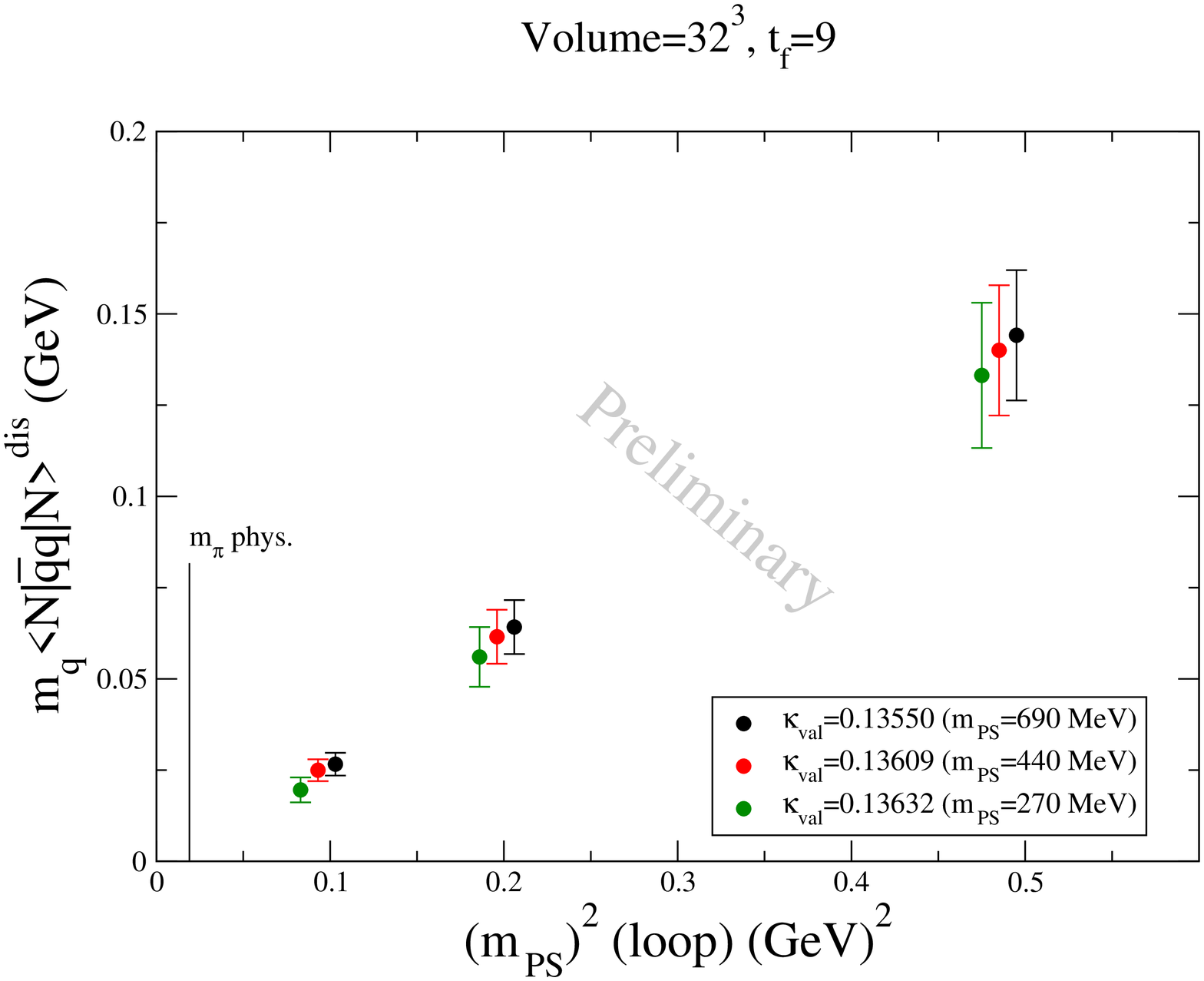}}}
\caption{Results for the scalar matrix element: (left) $f_{T_s}$ as a
  function of $m_{\mathrm{PS}}^2$ for the valence quark mass~(i.e. the mass of
  the quarks in the nucleon) for the three volumes
  studied. (right) $m_q\langle N|\bar{q}q|N\rangle^{\mathrm{dis}}$ as a
  function of $m_{\mathrm{PS}}^2$ for the loop quark mass on the $32^3\times
  64$ volume, for the three valence quark masses.}
\label{scalar}
\end{figure}

The results for $f_{T_s}$ are presented in fig.~\ref{scalar} as 
functions of $m_{\mathrm{PS}}^2$ corresponding to the mass of the~(valence)
quarks in the nucleon. No renormalization is required as the
combination, $m_q\langle N|\bar{q}q|N\rangle$, is scale and scheme
independent. There is consistency between the values obtained on the
different volumes for the heaviest valence quark mass, however, the
spread between the results increases when the quark mass is reduced.
Whether this is an indication of finite size effects will be clarified
once the statistics for the $40^3$ volume is increased.

In fig.~\ref{scalar} we also display the results for the disconnected
scalar matrix element, $m_q\langle N |\bar{q}q|N\rangle$, for the $32^3$
volume as a function of $m_{\mathrm{PS}}^2$ corresponding to the loop
quark mass. This combination is relevant for extracting the sigma
term,
\begin{equation}
\sigma_N = m_q\langle N|\bar{u}u+\bar{d}d|N\rangle\,.
\end{equation}
We found $2m_q\langle N|\bar{u}u|N\rangle^{\mathrm{dis}}$ for
$\kappa_{\mathrm{val}}=\kappa_{\mathrm{loop}}=\kappa_{\mathrm{sea}}=0.13632$ to amount to roughly $40\%$
of the connected contribution to $\sigma_N$ for this volume. However,
a more sophisticated method of extracting the matrix element from
$R^{\mathrm{dis}}$ is required in order to make a firm comparison.

\section{The spin contribution: $\Delta s$ and $\Delta q$}
\begin{figure}
\centerline{
\rotatebox{270}{\includegraphics[height=.48\textwidth,clip]{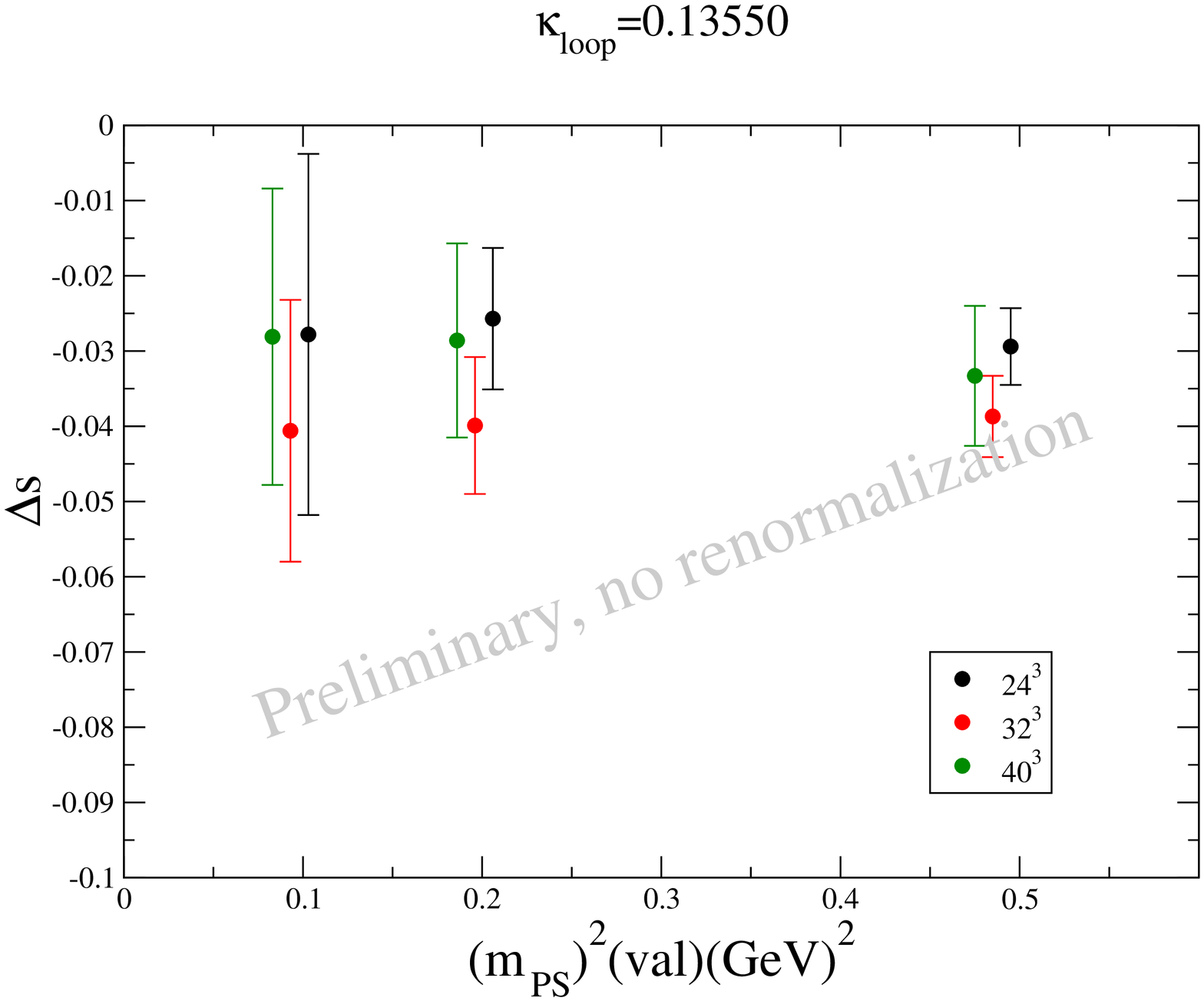}}\hspace*{.035\textwidth}
\rotatebox{270}{\includegraphics[height=.48\textwidth,clip]{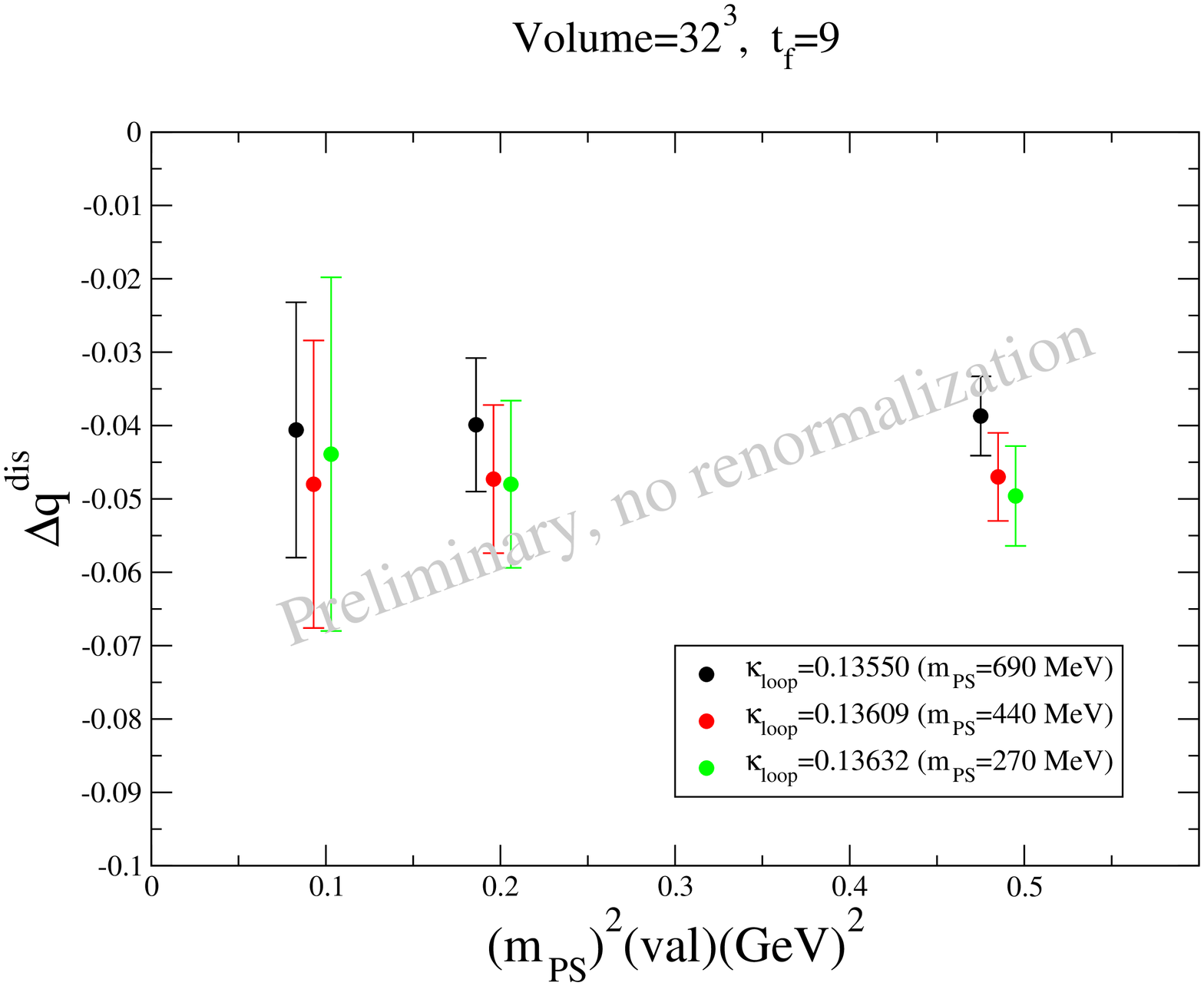}}
}
\caption{Results for $\Delta q$: (left) $\Delta s$ as a function of
  $m_{\mathrm{PS}}^2$ for the valence quark mass~(i.e. the mass of the quarks
  in the proton) for the three volumes studied. (right) $\Delta
  q^{\mathrm{dis}}$ from the $32^3\times 64$ volume for different loop quark
  masses, again as a function of $m_{\mathrm{PS}}^2$ for the valence quark mass.}
\label{deltaq}
\end{figure}

The results for $\Delta s$ on all three volumes are shown in
fig.~\ref{deltaq}.  No significant dependence on the valence quark
mass nor on the lattice size is seen in the data. Neither is there any
significant variation in the results if the loop quark mass is
reduced. These numbers will have to be multiplied by a renormalization
constant of approximately $0.8$ for the $\overline{MS}$
scheme~\cite{Skouroupathis:2008mf}. In contrast to the scalar case,
the disconnected contributions are much smaller than the connected
terms, at around $10\%$ for $\Delta d$ and $5\%$ for $\Delta u$.

\section{Outlook}
In the short term, the aim is to reach our target statistics of $2000$
trajectories for each volume. We then plan to begin an analysis close
to the physical sea quark mass. The nonperturbative renormalization
for $\Delta q$ needs to be calculated while for the scalar strangeness matrix
element mixing with the light flavours needs to be considered.

\acknowledgments This work was supported by the EU ITN STRONGnet, the
I3 HadronPhysics2 and the DFG SFB/Transregio 55.  Sara Collins
acknowledges support from the Claussen-Simon-Foundation (Stifterband
f\"ur die Deutsche Wissenschaft). Computations were performed on the
IBM BlueGene/L at EPCC (Edinburgh,UK), Regensburg's Athene HPC cluster,
the BlueGene/P (JuGene) and the Nehalem Cluster (JuRoPA) of the
J\"ulich Supercomputer Center and the SFB/TR55 QPACE supercomputers.
The Chroma software suite~\cite{Edwards:2004sx} was used extensively
in this work.

\end{document}